\documentclass[aps,prl,reprint,groupedaddress]{revtex4-1}

\usepackage{graphicx}
\usepackage{xcolor}

\begin{document}


\title{Photonic crystal carpet: Manipulating wave fronts in the near field \boris{at $1.55 \mu$m}}

\author{G. Scherrer$^{1}$, M. Hofman$^{2}$, W. {\'S}migaj$^{4,5}$, M. Kadic$^{3}$, T.-M. Chang$^{3}$, 
X. M{\'e}lique$^{2}$, D. Lippens$^{2}$, O. Vanb{\'e}sien$^{2}$, B. Cluzel$^{1}$, F. de Fornel$^{1}$, 
S. Guenneau$^{3}$, and B. Gralak$^{3}$\\[4mm]
\textit{$^{1}$ OCP--ICB, UMR CNRS 6303, 9 Avenue A. Savary, BP 47870, 21078 Dijon, France\\
$^{2}$ IEMN,UMR CNRS 8520, Universit{\'e} Lille 1, BP 60069, 59652 Villeneuve d'Ascq Cedex, France\\
$^{3}$ Institut Fresnel, UMR CNRS 7249, Universit{\'e} Aix-Marseille, Campus de St J{\'e}r{\^o}me, 
13397 Marseille Cedex 20, France\\
$^{4}$ Department of Mathematics, University College London, Gower Street, London WC1E 6BT, United Kingdom\\
$^{5}$ Faculty of Physics, Adam Mickiewicz University, Umultowska 85, 61-614 Pozna\'n, Poland\\
}}
\affiliation{}

\def\sebApril{}
\def\sebMay{}
\def\boris{}
\def\borisM{}
\def\seb{}
\def\wojtek{}

\date{\today}

\begin{abstract}
{\sebApril Ground-plane cloaks, which transform a curved mirror into a flat one,
and recently reported at wavelengths ranging from the optical to the visible spectrum,
bring the realm of optical illusion a step closer to reality. However, all carpet-cloaking experiments 
\wojtek{have thus far been carried out} in the far-field.}
{\sebApril Here,} we demonstrate numerically and experimentally that a dielectric photonic crystal (PC) 
of a complex shape made of a honeycomb array of air holes can scatter waves in the near field like a
{\sebApril PC with a flat boundary at stop band frequencies. This mirage effect relies upon a specific
arrangement of dielectric pillars placed at the nodes of a quasi-conformal grid dressing the PC.
Our carpet is shown to work throughout the range of wavelengths $1500\,$nm to $1650\,$nm
within the stop band \wojtek{extending from 1280 to 1940\,nm}}. The device \wojtek{has been} fabricated using a 
{single-mask advanced nanoelectronics technique} on III-V semiconductors and the near field measurements 
have been carried out in order to image the wave fronts{\sebApril 's} curvatures \sebMay{around
the telecommunication wavelength $1550\,$nm}.
\end{abstract}


\maketitle


\wojtek{The current interest in a better control of light through transformational
optics can be linked to the {\sebApril pioneering theoretical works of} 
Pendry et al.\ \cite{Pendry_2006} and Leonhardt
\cite{Leonhardt_2006} published in 2006.} The former seminal paper demonstrates the possibility of designing
a cloak that  renders any object inside it invisible to
electromagnetic radiation (using the covariant structure of Maxwell's equations),
while the latter concentrates on the ray\wojtek{-}optics limit (using conformal mappings in the
complex plane). In both cases, the cloak consists of a \wojtek{metamaterial}
whose physical properties (permittivity and permeability) are spatially varying and
matrix\wojtek{-}valued. These theoretical considerations might have remained an academic curiosity
{\sebApril without the experimental validation in 2006 \cite{Schurig_2006} of an electromagnetic
cloak making a copper cylinder invisible to an incident plane wave at $8.5\,$GHz. \wojtek{This demonstration sparked a
widespread interest in the optical community \cite{Kante_2009}.}

\wojtek{It must be noted that the originally proposed} invisibility cloak suffers from an inherent narrow bandwidth as
its transformational optics (TO) design leads to singular
(i.e.\wojtek{\ }infinitely anisotropic\wojtek{) permittivity and permeability} tensors on the
\wojtek{boundary} of the invisibility region. This is due to the fact that this invisibility region is created
by the blow\wojtek{-}up of a point in order to hide an object in the resulting metric hole. This TO approach to cloaking
therefore requires extreme material parameters, which can only be achieved with resonant structures.
This route to invisibility is reminiscent of the work of \sebMay{Greenleaf} et al.\wojtek{\ }in the context of electrical
impedance tomography \cite{Greenleaf_2003}. Interestingly, Leonhardt and Tyc \cite{Tyc_2009} have extended
TO to \wojtek{n}on-\sebMay{Euclidean} spaces, making use of the stereographic projection of an higher-dimensional sphere
onto the space where the cloak needs to be constructed. In this way, superluminal issues can be avoided
\cite{tyc_2011}. Other proposals of cloak singularity removal include blowing\wojtek{\ }up a small ball
\cite{kohn_2008} or a segment \cite{Jiang_2008} instead of a point,
but these remain theoretical proposals to this date.}

{\sebApril However, the carpet-cloaking scheme proposed by Li and Pendry in 2008 \cite{Li_2008} 
circumvents all these issues, since its TO design is based upon an easily implementable 
one-to-one map (a curved surface is transformed into a flat surface). In layman terms, the 
philosophy of this renewed approach to cloaking is to sweep dust under \sebMay{a} bump  
in a carpet. The challenge is that the bump should not be detected either by an external 
observer! One possible way to achieve this trick is to make a very small bump, but this
would have limited interest for applications \cite{cho_2010}. The bold idea of Li and Pendry
was to choose instead to create a mirage effect by curving wave trajectories
around the bump so that the carpet \sebMay{would appear} to be flat to an external observer.
Since then, carpet-cloaking has fuelled the experimental realization of cloaks at microwave,
terahertz, near-infrared and even visible frequencies 
\cite{Liu_2008,Lee_2009,Valentine_2009,Gabrielli_2009,Ergin_2010,Renger_2010}, \boris{including
a} carpet \wojtek{hiding} a macroscopic object in the visible spectrum \cite{chen_2011}.}

\begin{figure}
\resizebox{80mm}{!}{\includegraphics{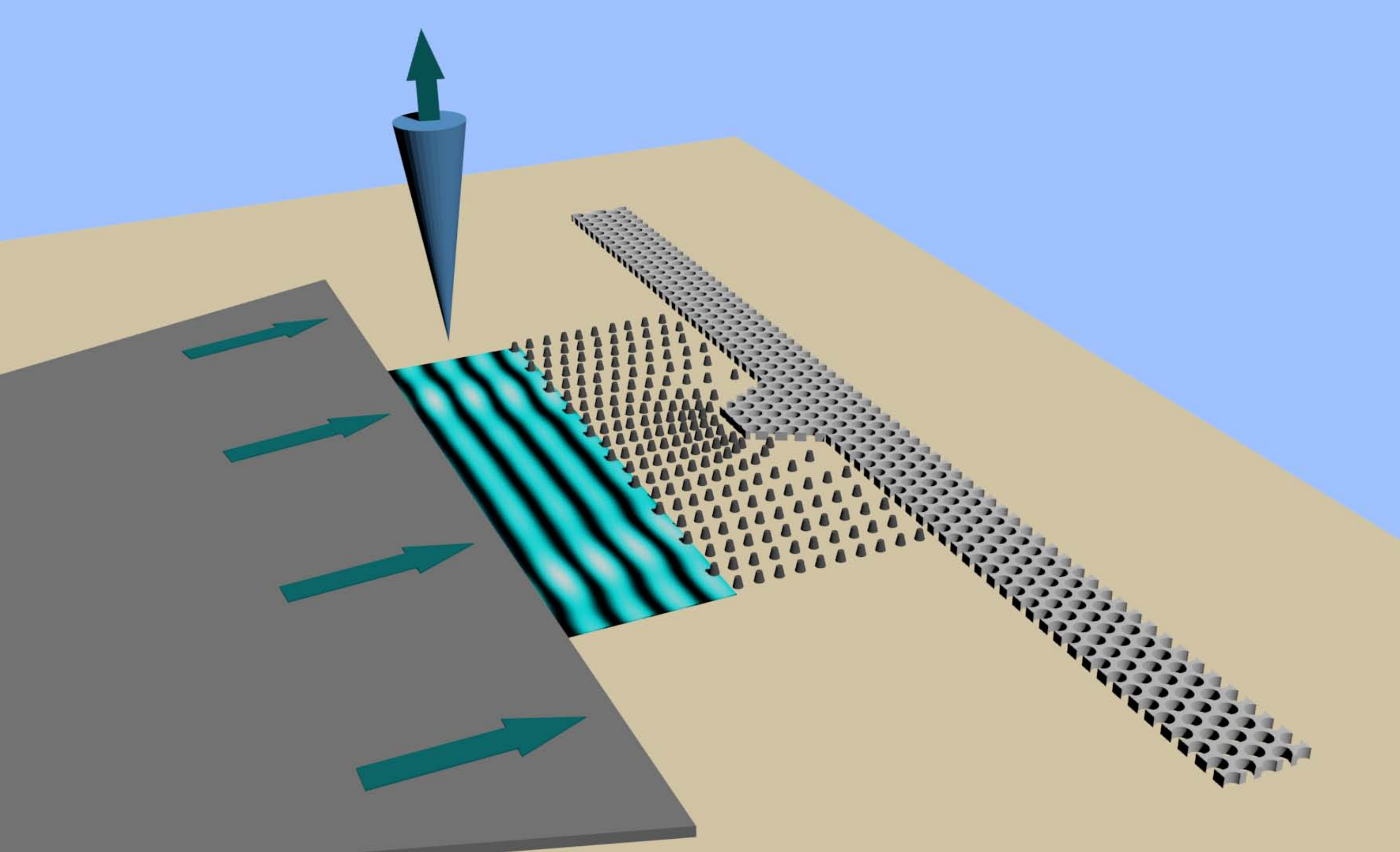}}
\caption{\sebMay{Experimental }\boris{ setup (from right to left): } \sebMay{F}lat mirror with 
\sebMay{a} bump of trapezoidal shape made of 
holes arranged \sebMay{along} a periodic array; \sebMay{S}urrounding carpet made of pillars\sebMay{; 
Injection guide generating an 
incident plane wave. Above the structure, \sebMay{p}robe measuring the near field.}\label{fig-1}}
\end{figure}

\wojtek{Herein, we propose} to mimic the optical response of a flat
mirror by a photonic crystal (PC) of irregular shape for wavelengths running in
its band gap. 
This optical illusion is achieved using 
\boris{dielectric} pillars with subwavelength \boris{diameter}. \boris{The ability of 
reconstructing the wave fronts in the \wojtek{vicinity} of the carpet is shown using near-field 
scanning optical microscopy (SNOM) \cite{Schonbrun_2007}.} 
At optical frequencies, the previous TO-based carpet-cloaks were demonstrated by far-field 
microscopy methods \cite{Liu_2008,Lee_2009,Valentine_2009} or leakage radiation microscopy \cite{Gabrielli_2009}. 
Those techniques are not suitable for 
our experimental setup described in Fig.\ 1. In this configuration, light successively propagates through 
free space and guiding areas over larges distances (more than 10\,$\mu$m between the very end of the injection 
guide and the input interface of the flat mirror). Thus, far-field microscopy would mainly allow access to 
scattered light such as out-of-plane losses, and leakage radiation microscopy would not be appropriate to 
measure wave fronts in the free space area. Such disadvantages clearly disappear with a near-field technique. 
SNOM provides a direct local probing of the electric field with a high spatial resolution thanks 
to the piezoelectric scanners associated with the probe. 
This markedly enhances our capabilities to
manipulate light, even in the extreme near field limit, when a \wojtek{source} lies in the close
neighbourhood of the carpet-cloak \cite{Zolla_2007}.

The plan of the paper is as follows. We first describe a PC which is used to mimic 
a mirror and to realize the reference structure. A trapezoidal bump 
\sebMay{on one side} of this PC plays the role of the object to hide. 
Next, a pillar structure designed using \sebMay{a} quasi\wojtek{-}conformal mapping is proposed to cloak 
the \sebMay{PC} bump. Numerical simulations have been performed to show the 
efficiency of the design for wavelengths ranging from {1500\,nm to 1650\,nm}. 
These \sebMay{composite} strutures consisting of a combination of \sebMay{air} holes 
and \sebMay{dielectric} pillars have been fabricated with a novel one-mask process.
Finally, near field measurements \seb{demonstrate} the ability of the carpet to hide the bump 
in the near field at frequencies around \sebMay{the telecommunication wavelength} 1550nm.

\begin{figure}
\resizebox{75mm}{!}{\includegraphics{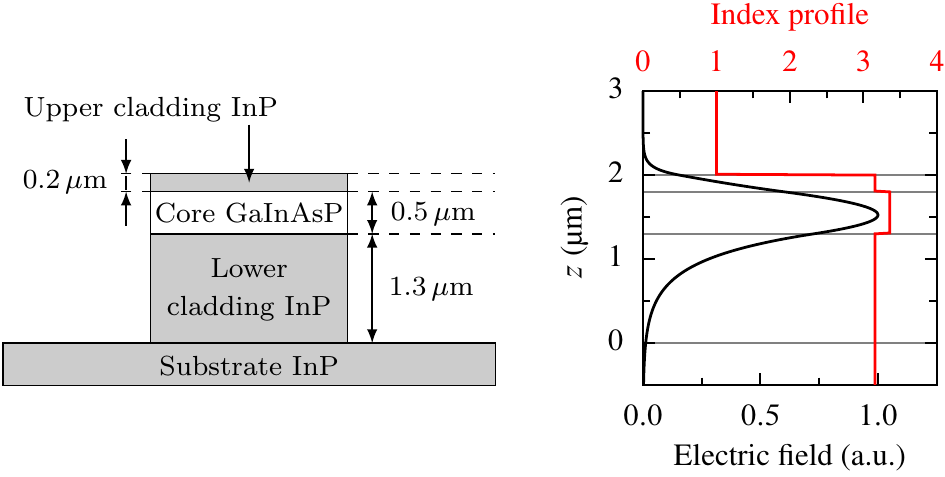}}
\caption{{Left: Epitaxial sequence of the guiding heterostructure. Right: Index profile and 
supported $p$-fundamental mode in the vertical direction.}\label{fig-2}}
\end{figure}

\begin{figure}[h]
\resizebox{75mm}{!}{\includegraphics{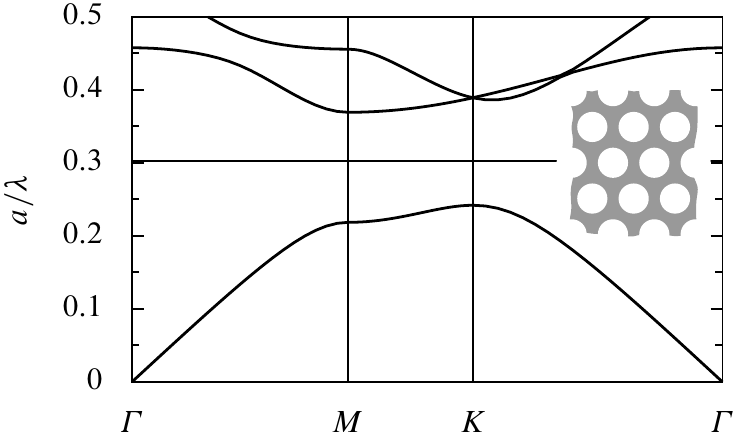}}
\caption{The band structure of the photonic crystal \seb{for the $p$-polarization}.
\borisM{The telecommunication wavelength 
1.55\,$\mu$m corresponds to $a / \lambda = 0.303$ \wojtek{(marked by horizontal line)}.}
\label{fig-3}}
\end{figure}

The first \sebMay{step} is to \sebMay{consider a PC consisting} of \sebMay{air} holes in a dielectric 
matrix to build \sebMay{an efficient} mirror \sebMay{at optical wavelengths}. In view of the fabrication, 
the guiding structure defined in 
\cite{Fabre_2008} is considered {(see Fig. \ref{fig-2})}. It is made of a InGaAsP guiding layer with thickness 
$500\text{nm}$ and index 3.36, sandwiched \wojtek{between a 200-nm thick layer of InP 
with index 3.16 and a substrate made of the same material. 
For the sake of simplicity, this multilayer is \sebMay{modelled} by \wojtek{an} effective medium of
 index 3.26, corresponding to the effective index of the fundamental $p$ mode of this heterostructure 
{(Fig. 2, right panel)}.}
\wojtek{As shown in Fig.\wojtek{\ }\ref{fig-3}, introduction of a hexagonal lattice of air holes of diameter 
$d = 347$\,nm and period $a = 470$\,nm into this homogeneous medium leads to the formation of a 
$p$-polarization band gap for wavelengths $\lambda$ ranging from 1280\,nm to 1940\,nm, i.e.\ for 
$a/\lambda$ from $0.242$ to $0.369$.}

\begin{figure}[h]
\resizebox{75mm}{!}{\includegraphics{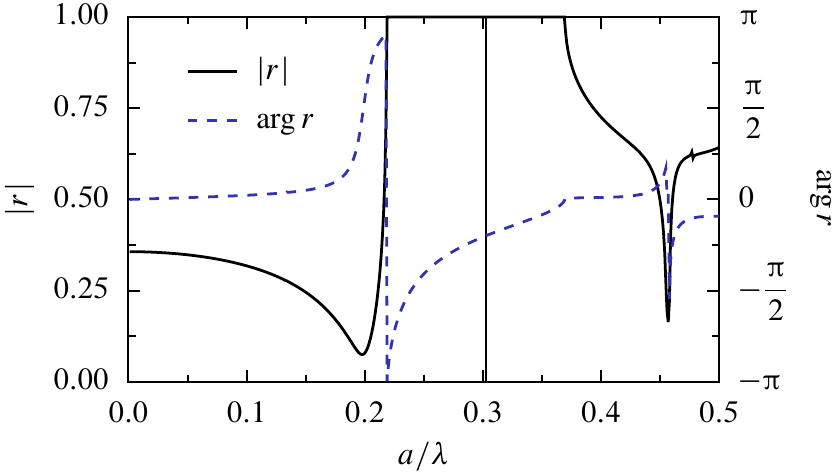}}
\caption{Variation of the reflection coefficient of the \sebMay{photonic} crystal \wojtek{at normal incidence}.
\sebMay{It is noted that} \wojtek{within the stop band $r$ has unit magnitude, but its phase varies monotonically from $-\pi$ to $0$.}
\label{fig-4}}
\end{figure}

\wojtek{An additional} challenge compared to a metallic mirror is the variation of the phase of the reflection
coefficient within the band gap. 
Indeed, compared to ground carpet cloaks designed for curved metal surfaces, where the boundary condition is 
of the Dirichlet or Neumann type depending upon light's polarization, a carpet cloak on a dielectric half-space 
\cite{Zhang_2008} requires greater care, since the mathematical model now involves (anisotropic) transmission 
conditions at the interface between the carpet and the dielectric half-space (which in our case consists of a PC).
In Fig.\wojtek{\ }\ref{fig-4}, the phase of reflection coefficient 
\wojtek{at} normal incidence is shown to vary \sebMay{monotonically} from $- \pi$ to zero \wojtek{across the stop 
band}. 
\borisM{The resulting} dispersion will \wojtek{have} an adverse \sebMay{e}ffect \wojtek{on} broadband 
cloaking. 
\sebMay{\borisM{More particularly,} it is noted that, at the telecommunication wavelength 
\boris{corresponding to} $a/\lambda=0.303$, the phase \borisM{$\arg(r)=-\wojtek{35}^\circ$} of the
PC is much different from the $\pi$ phase shift of a mirror, which is only achieved at the 
lower edge of the stop band}.
\borisM{This makes an unusual situation for the design of the carpet.}

The goal of carpet\wojtek{-}cloaking is to hide the bump of
a trapezoidal shape as shown in \wojtek{Fig.\ }\ref{fig-1} using an heterogeneous metamaterial 
(a medium with a subwavelength structure). In the present case, we opted for a carpet 
made of dielectric pillars. For this we build a quasi-grid 
deduced from conformal\wojtek{-}optics 
considerations \cite{Li_2008,Liu_2008,Valentine_2009,Gabrielli_2009,Ergin_2010,Renger_2010}: 
this approach preserves the
isotropy of the medium, unlike conventional \wojtek{TO} tools used for cloaking
\cite{Pendry_2006}.
We then place 
cylinders of diameter $200$nm at the nodes of the transformed grid.
We importantly note that the minimization of the modified Liao functional \cite{Liao_1993}, which 
is usually invoked in recent papers on the carpet-cloaking, can be replaced by a potential 
computation. 
In other words, we build a conformal grid which preserves the right angles. If one
can set up an electrostatic problem where the vertical lines are the electric field lines, 
then one can easily compute the equipotentials which are orthogonal at each point. The index 
distribution is then obtained numerically for each elementary 
cell of the grid using a simple vector product \cite{Li_2008}. 
The structure is then built by placing the \sebMay{dielectric} pillars at the nodes of the conformal grid.
\sebMay{We note that this approach has been successfully used for the design of a broadband plasmonic
carpet \cite{Renger_2010} and it is reminiscent of the work of Liang and Li \boris{\cite{Liang_2011}},
where 
Bloch waves follow curved trajectories in transformed photonic crystals. \wojtek{The
air holes in the PC and the dielectric pillars in the carpet have comparable diameters,}
hence the pillars are not deeply subwavelength in the stop band\wojtek{.} Our PC carpet \wojtek{can therefore} be
considered as an intermediate configuration between a transformed PC (Bragg regime) and
a conformal metamaterial (homogenization regime), keeping in mind artificial anisotropy is
unwanted here.}


\begin{figure}
\resizebox{75mm}{!}{\includegraphics{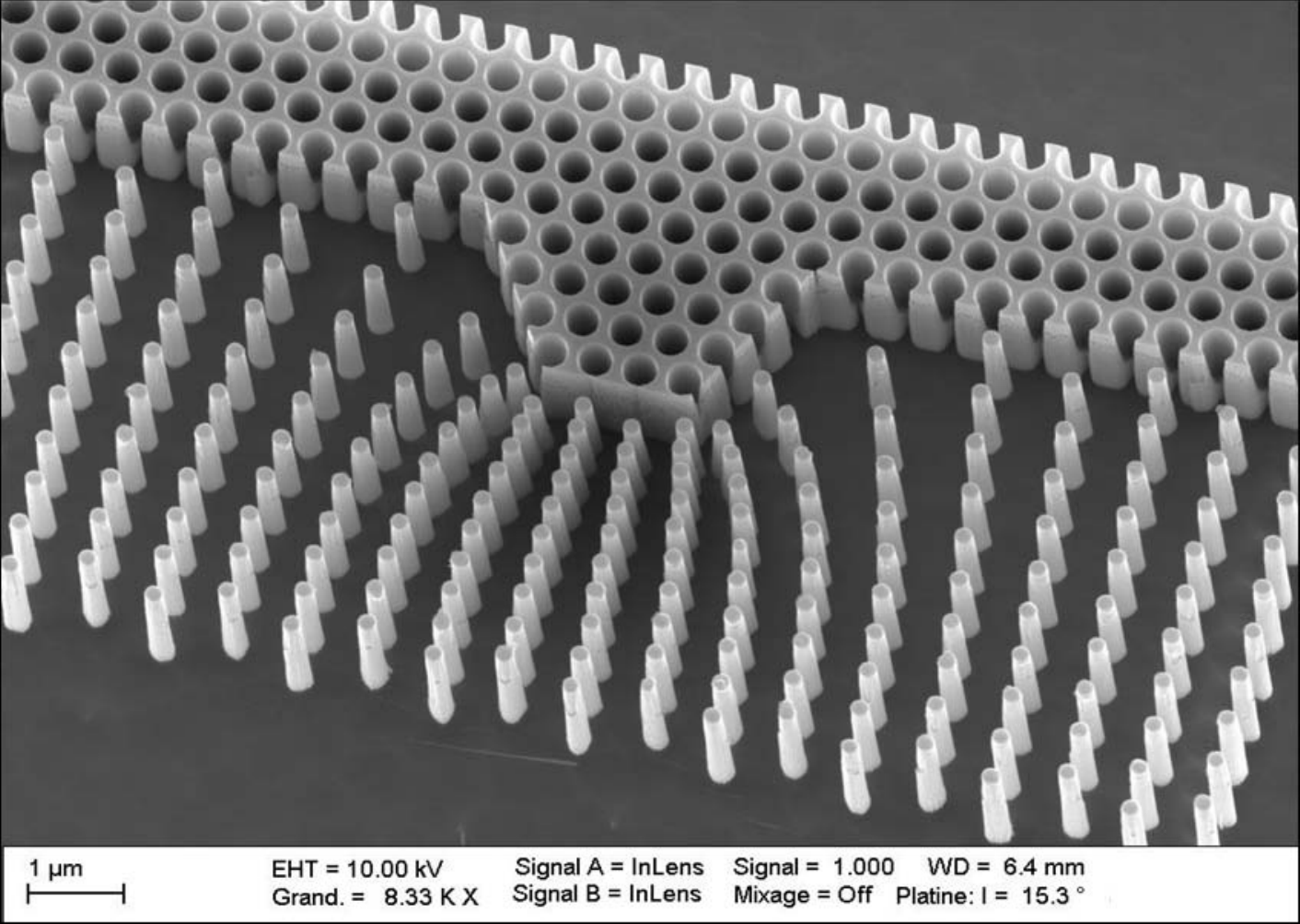}}
\caption{SEM (scanning electron microscope) picture of the InP-based fabricated device operating at 
$1.55 \,\mu\text{m}$: \sebMay{PC} reflector (period $470\, \text{nm}$, holes\sebMay{'} diameter 
$347 \,\text{nm}$) and \sebMay{dielectric PC carpet} (pillars\sebMay{'} diameter 
$200 \,\text{nm}$ and height $2 \,\mu\text{m}$).
\label{fig-5}}
\end{figure}

The device is fabricated using advanced nanoelectronics techniques on III--V semiconductors. 
We start from an InP--based heterostructure 
(InP--$200\,\text{nm}$/InGaAsP--$500\,\text{nm}$/InP--$1300\,\text{nm}$) 
grown by molecular beam epitaxy to confine light (at $1.55\, \mu \text{m}$) in the vertical direction. 
To create the two-dimensional patterning, we take benefit of the one-mask process primarily 
developed for photonic crystal based flat lenses \cite{Fabre_2008}. 
In brief, we employ a negative 
resist (HSQ) mask used for the electron beam lithography step to define simultaneously the 
matrix of the reflector, the pillars of the cloaking area as well as the injection guide 
constituted here of a monomode ridge waveguide progressively enlarged to form the quasi-plane 
wave requested to illuminate the device. After development and a densification by an O$_2$ plasma 
of the HSQ resist, the mask is kept for the inductively coupled plasma (ICP) etching step. The 
fabricated device, shown in \sebMay{Fig.}\ \ref{fig-5}, \wojtek{confirms that the process is able to} achieve in one step 
200 nm diameter pillars even in close proximity (separation less than 50 nm in front of the 
reflector) and a PC
with a period of 470\,nm and 347\,nm 
hole diameter over a depth larger than $1.5 \,\mu \text{m}$. The highly anisotropic character of the 
etching is preserved\sebMay{,} which permits to reproduce with a high degree of accuracy the theoretical 
dimensions targeted. 

The etching depth, larger than $1.5\, \mu \text{m}$, plays a vital role. 
{For our combined hole/pillar device, etching kinetics depends on the local environment due to species evacuation.  
Etching velocity is significantly lower in hole areas than in pillar regions. The process has been optimized to reach hole depths 
of about $1.5\, \mu \text{m}$
\cite{Fabre_2006}, leading to pillar heights between 2 and $2.2\, \mu \text{m}$. Both values are sufficient to ensure that} the 
integrated device \sebMay{works} in a regime which can be accurately modelled by a two-dimensional system 
since most of the electromagnetic energy of the guided mode will be \sebMay{controlled} by the etching\sebMay{, 
as demonstrated in Fig.\ \ref{fig-2}}. In addition, it can be expected \sebMay{from Fig.\ \ref{fig-2}} that the 
out\wojtek{-}of\wojtek{-}plane losses are not significant and do not 
disturb the behavior of the carpet\sebMay{, an assumption which will be experimentally validated in the sequel}.
This \sebMay{large etching depth is} a key point of {our fabrication process which combines holes and pillars: 
This} yields high quality 
results for the field measurements. 


\sebMay{Indeed, in Fig.\ \ref{fig-6} the} concept is verified using {SNOM in collection mode \cite{Fabre_2008}. 
Thanks to the configuration of both devices and experimental set up, a phase imaging is performed by a unique near 
field intensity measure without optical heterodyne detection \cite{Schonbrun_2007}.}
The measurement of the reflection on a flat photonic crystal (\sebMay{upper panel}) is similar to the
\sebMay{lower panel} (cloaked bump). We can see a significant improvement compared to the bump
\sebMay{taken} on its own \sebMay{in the middle panel}.
Before considering the optical images, it is important to note that the deep etching of the structures 
leads to large vertical displacements of the probe during the near-field scans\sebMay{, } which may affect the 
formation of the optical images. However, despite \sebMay{this} strongly modulated topology, the 
submicrometric features of the samples are properly resolved in all cases, which is particularly 
challenging 
on the sharp pillars of the cloak as well as on the photonic crystal mirrors. A three\wojtek{-}dimensional view 
of the measured topography is provided for illustration purposes 
in \sebMay{Fig.\ \ref{fig-8}a}.
\begin{widetext}
\begin{center}
\begin{figure}[!h]
\resizebox{150mm}{!}{\includegraphics{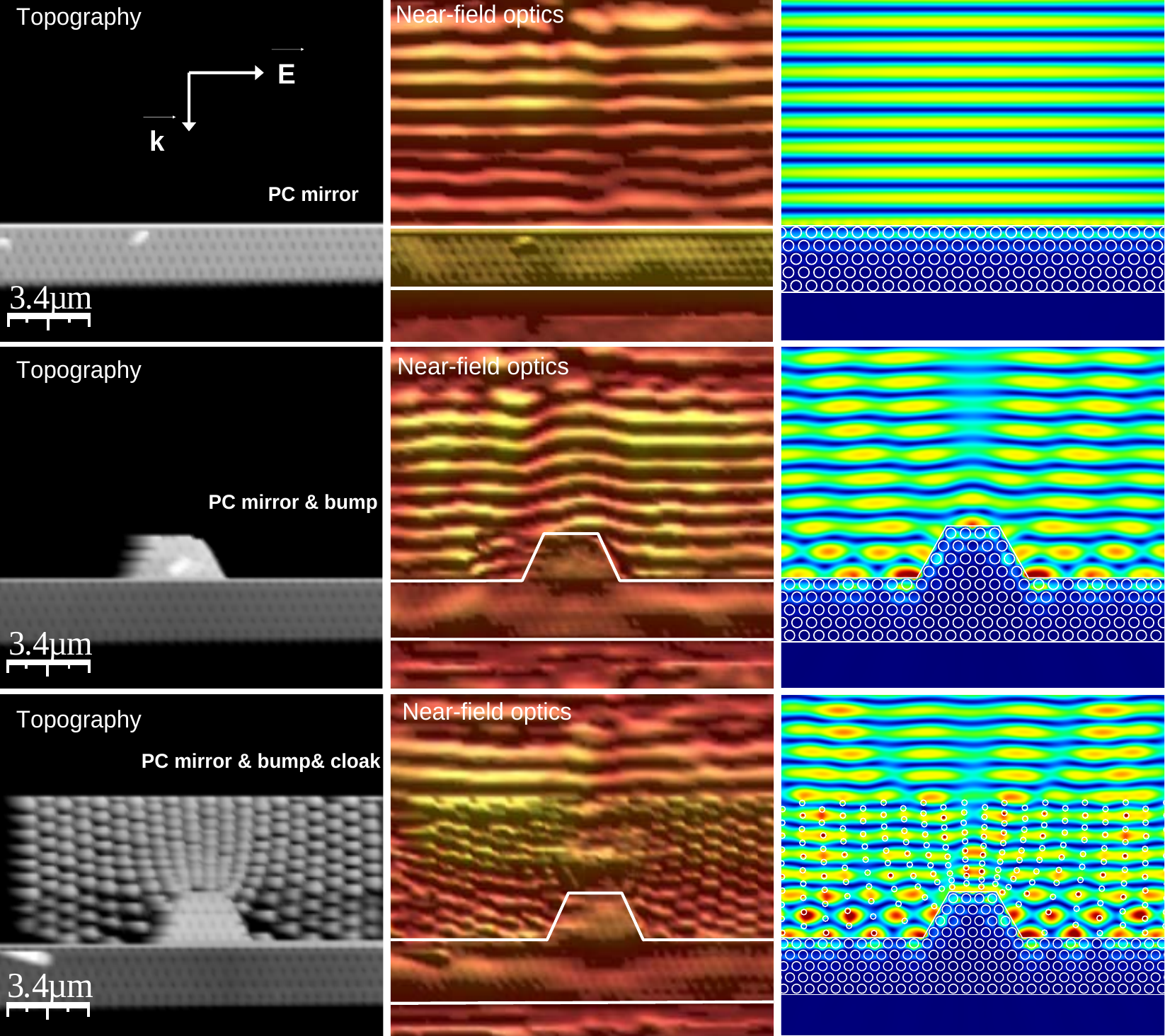}}
\caption{\boris{Left column:}
Top view of the topographical images recorded by SNOM above the two reference structures, 
the flat mirror and the mirror with bump, and above the cloaked structure. \boris{Central column:} Corresponding 
optical near-field {intensity} images recorded for a 1540nm wavelength. 
\seb{Right column: Numerical simulations for the {amplitude} of the field {from the blue for 
the null amplitude to the red for the maximum (see the color bar on Fig. \ref{fig-9} for an incident plane 
wave with amplitude set to unity)}.}
\label{fig-6}}
\end{figure}
\end{center}
\end{widetext}
	
The optical images, shown in \sebMay{Fig.\ \ref{fig-6}}, present the typical optical behaviour of the different structures 
at \sebMay{a} telecommunication wavelength of \sebMay{1540\,nm}. A standard illuminated view of the optical images is 
chosen here for the images representation because of a lateral modulation of the intensity of the plane wave impinging 
on the structure\sebMay{. This can be} attributed to a non\wojtek{-}adiabatic tapering 
of the input waveguides. 
However, as shown hereafter, this is not an issue for the reported observations since it is the shape 
of the wave fronts that are required for interpretation and not their relative intensity.

On the optical images, one can observe the standing wave patterns between the incident and reflected 
wave by the photonic crystal mirror. In the case of the reference structure made of the \sebMay{PC} 
flat mirror, flat interference fringes with a half-wavelength spatial periodicity are clearly resolved 
on the optical near-field images. Then, introducing a defect on the planarity of \sebMay{the PC}
mirror  surface, namely \sebMay{a} bump, leads to a local deformation of the planarity of the standing wave pattern. 
\borisM{More precisely,} as clearly visible on the near-field measurement, a half\wojtek{-}period shift of the 
fringes is measured locally above the bump  on the mirror, which corresponds to a local phase shift 
of the reflected wave. As a consequence, cloaking the bump would result in suppressing this phase 
shift on the observed interference pattern. Experimentally, the near-field images recorded on the 
cloak structures \wojtek{reveal} almost flat fringes \wojtek{in front of} the cloak\sebMay{, } 
which \wojtek{testifies} that the bump has been 
properly cloaked. 
\begin{figure}[b]
\resizebox{65mm}{!}{\includegraphics{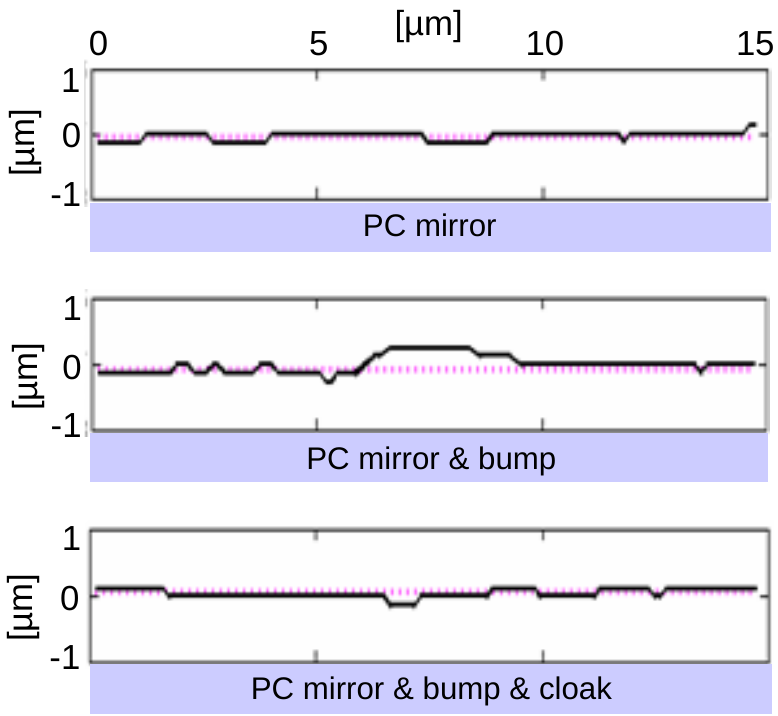}}
\caption{\borisM{Lines following the maxima of the field intensity for the mirror (top), 
the mirror with bump (middle) and the cloaked structure (bottom).
\seb{The resulting reduction of the bump signature is 4.75 
(computed using the blue curve in the upper panel as a benchmark, and comparing its area against 
that of the blue curves in the middle and lower panels).}
}\label{fig-7}}
\end{figure}
Since the contrast of these fringes remains the same \sebMay{as} in the reference cases 
without cloak and since the reflectivity of the first row of the cloak pillars is low due to their 
subwavelength size, these flat fringes clearly result from the interference between the wave reflected 
by the photonic crystal mirror after passing through the cloak and the incident plane wave. Then, inside 
the cloak, as visible on all the presented images in \sebMay{Fig.\ \ref{fig-6} and \ref{fig-8}}, the measured optical 
near-field distributions are unfortunately highly modulated by the cloak topography (\sebMay{Fig.\ \ref{fig-8}a}), 
making us blind 
to the standing wave pattern inside it.

\borisM{In addition, a quantitative \wojtek{measure} of the cloaking \wojtek{efficiency} can be derived from the evaluation of the
modif\wojtek{i}ed Liao functional \cite{Liao_1993}. 
The fringes are defined as the lines corresponding to the 
maxima of the field intensity. Figure \ref{fig-7} shows these fringes for the three situations. 
According to the 
method of least squares, it is found that the deviations from a $15\,\mu\text{m}$ segment are respectively 37, 129 and 
52 for the flat mirror, the flat mirror with bump, and the cloaked structure (arbitrary unit is used).
These raw data show that the measurements and the device's imperfections induce uncertainties which make the error 
bars very wide.
However, when restricting the evaluation of the deviation to the $5\,\mu\text{m}$ segment centred above the bump, 
the deviations become respectively 13, 76 and 16. In this case, it is found that the carpet leads to a reduction 
of the bump signature by a factor of 4.75\seb{, which is comparatively higher than the reduction factor of 3.7 obtained
for a plasmonic carpet at 800\,nm in \cite{Renger_2010}.} 
}

\begin{figure}[t]
\resizebox{80mm}{!}{\includegraphics{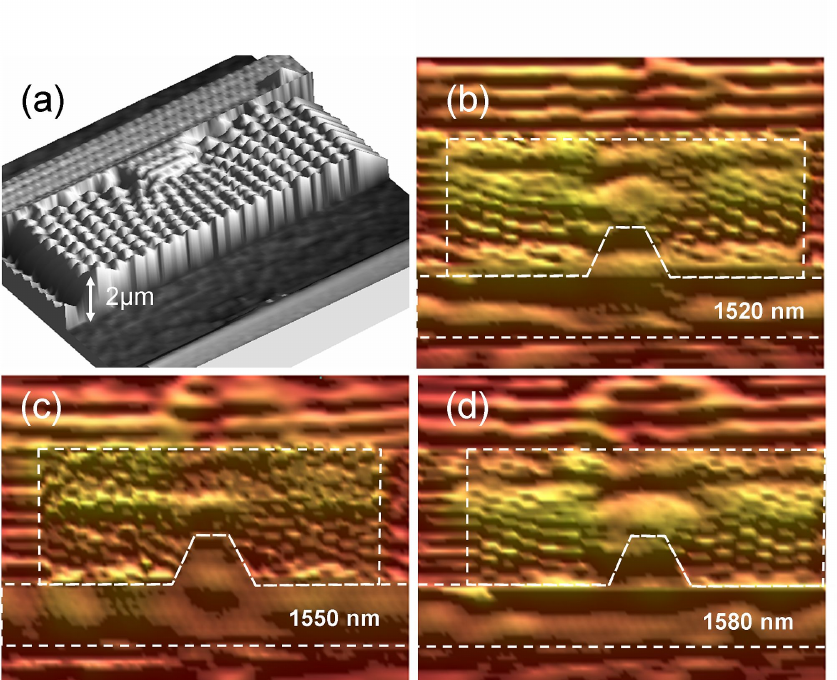}}
\caption{Three dimensional view \seb{of} topographical image\seb{s} of the cloak, bump and mirror recorded in near-field 
by SNOM. Corresponding near-field optical images recorded at (b) 1520, (c) 1550 and (d) 1580\,nm.\label{fig-8}}
\end{figure}

\begin{figure}
\resizebox{85mm}{!}{\includegraphics{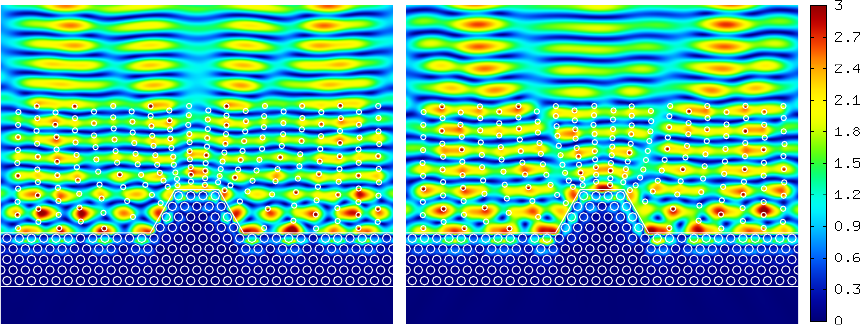}}
\caption{Numerical simulations for the {amplitude of the magnetic} field at 
$1500\,\text{nm}$ (left) and $1650\,\text{nm}$ (right).
{The color bar corresponds to an incident plane 
wave with amplitude set to unity}.
\label{fig-9}}
\end{figure}

As shown in Fig.\ \ref{fig-8}, such experimental features, the strongly topographically modulated light inside the 
cloak as well as the standing wave with flat fringes showing the cloak effect, are observed all over 
the C-band of telecommunication wavelengths, i.e. for wavelengths ranging from 1520\,nm to 1580\,nm. This 
wavelength range is limited experimentally at shorter wavelengths by the absorption of the InGaAs 
layer used for waveguiding and at longer wavelengths by our laser source. However, numerical \boris{simulations
show that the carpet operates for a broad band of wavelengths ranging from $1500\,\text{nm}$ to $1650\,\text{nm}$ 
(see figure \ref{fig-9}). 
}

In conclusion, a new cloaking device with a photonic crystal mirror has been designed to 
operate at the telecommunication wavelength 1550\,nm. This device has been fabricated using an original
technique based on a novel one-mask process which makes it possible to create pillars and
holes with an etching depth larger than $1.5\,\mu\text{m}$. The characterisation of the cloaking structure 
has been performed with near-field measurements showing the interference 
fringes in the closeness of the cloaking structure. The \seb{invisibility} carpet is shown to operate 
\seb{very efficiently} for wavelengths ranging from 1520\,nm to 1580\,nm (and from $1500$\,nm to $1650$\,nm in theory): 
A reduction factor of 4.75 has been achieved for the bump signature. We note that our design
should also work for visible wavelengths: To achieve this goal theoretically, 
one need only scale 
down the PC structure by a factor 2, as in this case, the stop band ranges from 670\,nm to 1020\,nm 
for a lattice constant of 235\,nm and a diameter of air holes of 173.5\,nm.
The main difficulty \wojtek{lies in fact in downscaling the carpet cloak}, as this would bring down 
the diameter of pillars\wojtek{\ }to 100\wojtek{\,}nm, which remains a technological challenge for us: 
The pillars should be tall enough in order to ensure the two-dimensional nature of the problem.

\begin{acknowledgments}
This work was partly supported by the project FANI
funded by the Agence Nationale de la Recherche (ANR).
S.G. acknowledges funding from the European Union through ERC Starting Grant Anamorphism.
G.S. would like to thank the Regional Council of Burgundy for financial support.
\boris{Dr.} \sebMay{Jensen} \boris{Li is gratefully acknowledged for useful discussions.}
\end{acknowledgments}


\begin{thebibliography}{25}%
\makeatletter
\providecommand \@ifxundefined [1]{%
 \@ifx{#1\undefined}
}%
\providecommand \@ifnum [1]{%
 \ifnum #1\expandafter \@firstoftwo
 \else \expandafter \@secondoftwo
 \fi
}%
\providecommand \@ifx [1]{%
 \ifx #1\expandafter \@firstoftwo
 \else \expandafter \@secondoftwo
 \fi
}%
\providecommand \natexlab [1]{#1}%
\providecommand \enquote  [1]{``#1''}%
\providecommand \bibnamefont  [1]{#1}%
\providecommand \bibfnamefont [1]{#1}%
\providecommand \citenamefont [1]{#1}%
\providecommand \href@noop [0]{\@secondoftwo}%
\providecommand \href [0]{\begingroup \@sanitize@url \@href}%
\providecommand \@href[1]{\@@startlink{#1}\@@href}%
\providecommand \@@href[1]{\endgroup#1\@@endlink}%
\providecommand \@sanitize@url [0]{\catcode `\\12\catcode `\$12\catcode
  `\&12\catcode `\#12\catcode `\^12\catcode `\_12\catcode `\%12\relax}%
\providecommand \@@startlink[1]{}%
\providecommand \@@endlink[0]{}%
\providecommand \url  [0]{\begingroup\@sanitize@url \@url }%
\providecommand \@url [1]{\endgroup\@href {#1}{\urlprefix }}%
\providecommand \urlprefix  [0]{URL }%
\providecommand \Eprint [0]{\href }%
\providecommand \doibase [0]{http://dx.doi.org/}%
\providecommand \selectlanguage [0]{\@gobble}%
\providecommand \bibinfo  [0]{\@secondoftwo}%
\providecommand \bibfield  [0]{\@secondoftwo}%
\providecommand \translation [1]{[#1]}%
\providecommand \BibitemOpen [0]{}%
\providecommand \bibitemStop [0]{}%
\providecommand \bibitemNoStop [0]{.\EOS\space}%
\providecommand \EOS [0]{\spacefactor3000\relax}%
\providecommand \BibitemShut  [1]{\csname bibitem#1\endcsname}%
\let\auto@bib@innerbib\@empty
\bibitem [{\citenamefont {Pendry}\ \emph {et~al.}(2006)\citenamefont {Pendry},
  \citenamefont {Schurig},\ and\ \citenamefont {Smith}}]{Pendry_2006}%
  \BibitemOpen
  \bibfield  {author} {\bibinfo {author} {\bibfnamefont {J.~B.}\ \bibnamefont
  {Pendry}}, \bibinfo {author} {\bibfnamefont {D.}~\bibnamefont {Schurig}}, \
  and\ \bibinfo {author} {\bibfnamefont {D.~R.}\ \bibnamefont {Smith}},\
  }\href@noop {} {\bibfield  {journal} {\bibinfo  {journal} {Science}\ }\textbf
  {\bibinfo {volume} {312}},\ \bibinfo {pages} {1780} (\bibinfo {year}
  {2006})} 
\bibitem [{\citenamefont {Leonhardt}(2006)}]{Leonhardt_2006}%
  \BibitemOpen
  \bibfield  {author} {\bibinfo {author} {\bibfnamefont {U.}~\bibnamefont
  {Leonhardt}},\ }\href@noop {} {\bibfield  {journal} {\bibinfo  {journal}
  {Science}\ }\textbf {\bibinfo {volume} {312}},\ \bibinfo {pages} {1777}
  (\bibinfo {year} {2006})} 
\bibitem [{\citenamefont {Schurig}\ \emph {et~al.}(2006)\citenamefont
  {Schurig}, \citenamefont {Mock}, \citenamefont {Justice}, \citenamefont
  {Cummer}, \citenamefont {Pendry}, \citenamefont {Starr},\ and\ \citenamefont
  {Smith}}]{Schurig_2006}%
  \BibitemOpen
  \bibfield  {author} {\bibinfo {author} {\bibfnamefont {D.}~\bibnamefont
  {Schurig}}, \bibinfo {author} {\bibfnamefont {J.~J.}\ \bibnamefont {Mock}},
  \bibinfo {author} {\bibfnamefont {B.~J.}\ \bibnamefont {Justice}}, \bibinfo
  {author} {\bibfnamefont {C.~A.}\ \bibnamefont {Cummer}}, \bibinfo {author}
  {\bibfnamefont {J.~B.}\ \bibnamefont {Pendry}}, \bibinfo {author}
  {\bibfnamefont {A.~F.}\ \bibnamefont {Starr}}, \ and\ \bibinfo {author}
  {\bibfnamefont {D.~R.}\ \bibnamefont {Smith}},\ }\href@noop {} {\bibfield
  {journal} {\bibinfo  {journal} {Science}\ }\textbf {\bibinfo {volume}
  {314}},\ \bibinfo {pages} {977} (\bibinfo {year} {2006})}
\bibitem [{\citenamefont {Kant{\'e}}\ \emph {et~al.}(2009)\citenamefont
  {Kant{\'e}}, \citenamefont {Germain}, ,\ and\ \citenamefont
  {de~Lustrac}}]{Kante_2009}%
  \BibitemOpen
  \bibfield  {author} {\bibinfo {author} {\bibfnamefont {B.}~\bibnamefont
  {Kant{\'e}}}, \bibinfo {author} {\bibfnamefont {D.}~\bibnamefont {Germain}},
  , \ and\ \bibinfo {author} {\bibfnamefont {A.}~\bibnamefont {de~Lustrac}},\
  }\href@noop {} {\bibfield  {journal} {\bibinfo  {journal} {Phys. Rev. B}\
  }\textbf {\bibinfo {volume} {80}},\ \bibinfo {pages} {201104} (\bibinfo
  {year} {2009})} 
\bibitem [{\citenamefont {Greenleaf}\ \emph {et~al.}(2003)\citenamefont
  {Greenleaf}, \citenamefont {Lassas},\ and\ \citenamefont
  {Uhlmann}}]{Greenleaf_2003}%
  \BibitemOpen
  \bibfield  {author} {\bibinfo {author} {\bibfnamefont {A.}~\bibnamefont
  {Greenleaf}}, \bibinfo {author} {\bibfnamefont {A.}~\bibnamefont {Lassas}}, \
  and\ \bibinfo {author} {\bibfnamefont {G.}~\bibnamefont {Uhlmann}},\
  }\href@noop {} {\bibfield  {journal} {\bibinfo  {journal} {Math. Res. Lett.}\
  }\textbf {\bibinfo {volume} {10}},\ \bibinfo {pages} {685} (\bibinfo {year}
  {2003})} 
\bibitem [{\citenamefont {Leonhardt}\ and\ \citenamefont
  {Tyc}(2009)}]{Tyc_2009}%
  \BibitemOpen
  \bibfield  {author} {\bibinfo {author} {\bibfnamefont {U.}~\bibnamefont
  {Leonhardt}}\ and\ \bibinfo {author} {\bibfnamefont {T.}~\bibnamefont
  {Tyc}},\ }\href@noop {} {\bibfield  {journal} {\bibinfo  {journal} {Science}\
  }\textbf {\bibinfo {volume} {323}},\ \bibinfo {pages} {110} (\bibinfo {year}
  {2009})} 
\bibitem [{\citenamefont {Percze}\ \emph {et~al.}(2011)\citenamefont {Percze},
  \citenamefont {Tyc},\ and\ \citenamefont {Leonhardt}}]{tyc_2011}%
  \BibitemOpen
  \bibfield  {author} {\bibinfo {author} {\bibfnamefont {J.}~\bibnamefont
  {Percze}}, \bibinfo {author} {\bibfnamefont {T.}~\bibnamefont {Tyc}}, \ and\
  \bibinfo {author} {\bibfnamefont {U.}~\bibnamefont {Leonhardt}},\ }\href@noop
  {} {\bibfield  {journal} {\bibinfo  {journal} {New Journal of Physics}\
  }\textbf {\bibinfo {volume} {13}},\ \bibinfo {pages} {083007} (\bibinfo
  {year} {2011})} 
\bibitem [{\citenamefont {Kohn}\ \emph {et~al.}(2011)\citenamefont {Kohn},
  \citenamefont {Shen}, \citenamefont {Vogelius},\ and\ \citenamefont
  {Weinstein}}]{kohn_2008}%
  \BibitemOpen
  \bibfield  {author} {\bibinfo {author} {\bibfnamefont {R.~V.}\ \bibnamefont
  {Kohn}}, \bibinfo {author} {\bibfnamefont {H.}~\bibnamefont {Shen}}, \bibinfo
  {author} {\bibfnamefont {M.~S.}\ \bibnamefont {Vogelius}}, \ and\ \bibinfo
  {author} {\bibfnamefont {M.~I.}\ \bibnamefont {Weinstein}},\ }\href@noop {}
  {\bibfield  {journal} {\bibinfo  {journal} {Inverse Problems}\ }\textbf
  {\bibinfo {volume} {24}},\ \bibinfo {pages} {015016} (\bibinfo {year}
  {2011})} 
\bibitem [{\citenamefont {Jiang}\ \emph {et~al.}(2008)\citenamefont {Jiang},
  \citenamefont {Cui}, \citenamefont {Yang}, \citenamefont {Cheng},
  \citenamefont {Liu},\ and\ \citenamefont {Smith}}]{Jiang_2008}%
  \BibitemOpen
  \bibfield  {author} {\bibinfo {author} {\bibfnamefont {W.~X.}\ \bibnamefont
  {Jiang}}, \bibinfo {author} {\bibfnamefont {T.~J.}\ \bibnamefont {Cui}},
  \bibinfo {author} {\bibfnamefont {X.~M.}\ \bibnamefont {Yang}}, \bibinfo
  {author} {\bibfnamefont {Q.}~\bibnamefont {Cheng}}, \bibinfo {author}
  {\bibfnamefont {R.}~\bibnamefont {Liu}}, \ and\ \bibinfo {author}
  {\bibfnamefont {D.~R.}\ \bibnamefont {Smith}},\ }\href@noop {} {\bibfield
  {journal} {\bibinfo  {journal} {Appl. Phys. Lett.}\ }\textbf {\bibinfo
  {volume} {92}},\ \bibinfo {pages} {264101} (\bibinfo {year}
  {2008})} 
\bibitem [{\citenamefont {Li}\ and\ \citenamefont {Pendry}(2008)}]{Li_2008}%
  \BibitemOpen
  \bibfield  {author} {\bibinfo {author} {\bibfnamefont {J.}~\bibnamefont
  {Li}}\ and\ \bibinfo {author} {\bibfnamefont {J.~B.}\ \bibnamefont
  {Pendry}},\ }\href@noop {} {\bibfield  {journal} {\bibinfo  {journal} {Phys.
  Rev. Lett.}\ }\textbf {\bibinfo {volume} {101}},\ \bibinfo {pages} {203901}
  (\bibinfo {year} {2008})} 
\bibitem [{\citenamefont {Cho}(2010)}]{cho_2010}%
  \BibitemOpen
  \bibfield  {author} {\bibinfo {author} {\bibfnamefont {A.}~\bibnamefont
  {Cho}},\ }\href@noop {} {\bibfield  {journal} {\bibinfo  {journal} {Science}\
  }\textbf {\bibinfo {volume} {329}},\ \bibinfo {pages} {277} (\bibinfo {year}
  {2010})} 
\bibitem [{\citenamefont {Liu}\ \emph {et~al.}(2008)\citenamefont {Liu},
  \citenamefont {Ji}, \citenamefont {Mock}, \citenamefont {Chin}, \citenamefont
  {Cui},\ and\ \citenamefont {Smith}}]{Liu_2008}%
  \BibitemOpen
  \bibfield  {author} {\bibinfo {author} {\bibfnamefont {R.}~\bibnamefont
  {Liu}}, \bibinfo {author} {\bibfnamefont {C.}~\bibnamefont {Ji}}, \bibinfo
  {author} {\bibfnamefont {J.~J.}\ \bibnamefont {Mock}}, \bibinfo {author}
  {\bibfnamefont {J.}~\bibnamefont {Chin}}, \bibinfo {author} {\bibfnamefont
  {T.~J.}\ \bibnamefont {Cui}}, \ and\ \bibinfo {author} {\bibfnamefont
  {D.~R.}\ \bibnamefont {Smith}},\ }\href@noop {} {\bibfield  {journal}
  {\bibinfo  {journal} {Science}\ }\textbf {\bibinfo {volume} {323}},\ \bibinfo
  {pages} {366} (\bibinfo {year} {2008})} 
\bibitem [{\citenamefont {Lee}\ \emph {et~al.}(2009)\citenamefont {Lee},
  \citenamefont {Blair}, \citenamefont {Tamma}, \citenamefont {Wu},
  \citenamefont {Rhee}, \citenamefont {Summers},\ and\ \citenamefont
  {Park}}]{Lee_2009}%
  \BibitemOpen
  \bibfield  {author} {\bibinfo {author} {\bibfnamefont {J.~H.}\ \bibnamefont
  {Lee}}, \bibinfo {author} {\bibfnamefont {J.}~\bibnamefont {Blair}}, \bibinfo
  {author} {\bibfnamefont {V.~A.}\ \bibnamefont {Tamma}}, \bibinfo {author}
  {\bibfnamefont {Q.}~\bibnamefont {Wu}}, \bibinfo {author} {\bibfnamefont
  {S.~J.}\ \bibnamefont {Rhee}}, \bibinfo {author} {\bibfnamefont {C.~J.}\
  \bibnamefont {Summers}}, \ and\ \bibinfo {author} {\bibfnamefont
  {W.}~\bibnamefont {Park}},\ }\href@noop {} {\bibfield  {journal} {\bibinfo
  {journal} {Optics Express}\ }\textbf {\bibinfo {volume} {17}},\ \bibinfo
  {pages} {12922} (\bibinfo {year} {2009})} 
\bibitem [{\citenamefont {Valentine}\ \emph {et~al.}(2009)\citenamefont
  {Valentine}, \citenamefont {Li}, \citenamefont {Zentgraf}, \citenamefont
  {Bartal},\ and\ \citenamefont {Zhang}}]{Valentine_2009}%
  \BibitemOpen
  \bibfield  {author} {\bibinfo {author} {\bibfnamefont {J.}~\bibnamefont
  {Valentine}}, \bibinfo {author} {\bibfnamefont {J.}~\bibnamefont {Li}},
  \bibinfo {author} {\bibfnamefont {T.}~\bibnamefont {Zentgraf}}, \bibinfo
  {author} {\bibfnamefont {G.}~\bibnamefont {Bartal}}, \ and\ \bibinfo {author}
  {\bibfnamefont {X.}~\bibnamefont {Zhang}},\ }\href@noop {} {\bibfield
  {journal} {\bibinfo  {journal} {Nature Mater.}\ }\textbf {\bibinfo {volume}
  {8}},\ \bibinfo {pages} {569} (\bibinfo {year} {2009})} 
\bibitem [{\citenamefont {Gabrielli}\ \emph {et~al.}(2009)\citenamefont
  {Gabrielli}, \citenamefont {Cardenas}, \citenamefont {Poitras},\ and\
  \citenamefont {Lipson}}]{Gabrielli_2009}%
  \BibitemOpen
  \bibfield  {author} {\bibinfo {author} {\bibfnamefont {L.~H.}\ \bibnamefont
  {Gabrielli}}, \bibinfo {author} {\bibfnamefont {J.}~\bibnamefont {Cardenas}},
  \bibinfo {author} {\bibfnamefont {C.~B.}\ \bibnamefont {Poitras}}, \ and\
  \bibinfo {author} {\bibfnamefont {M.}~\bibnamefont {Lipson}},\ }\href@noop {}
  {\bibfield  {journal} {\bibinfo  {journal} {Nat. Photonics}\ }\textbf
  {\bibinfo {volume} {8}},\ \bibinfo {pages} {461} (\bibinfo {year}
  {2009})} 
\bibitem [{\citenamefont {Ergin}\ \emph {et~al.}(2010)\citenamefont {Ergin},
  \citenamefont {Brenner}, \citenamefont {Pendry},\ and\ \citenamefont
  {Wegener}}]{Ergin_2010}%
  \BibitemOpen
  \bibfield  {author} {\bibinfo {author} {\bibfnamefont {N.}~\bibnamefont
  {Ergin}, \bibfnamefont {T.~andStenger}}, \bibinfo {author} {\bibfnamefont
  {P.}~\bibnamefont {Brenner}}, \bibinfo {author} {\bibfnamefont {J.~B.}\
  \bibnamefont {Pendry}}, \ and\ \bibinfo {author} {\bibfnamefont
  {M.}~\bibnamefont {Wegener}},\ }\href@noop {} {\bibfield  {journal} {\bibinfo
   {journal} {Science}\ }\textbf {\bibinfo {volume} {328}},\ \bibinfo {pages}
  {337} (\bibinfo {year} {2010})} 
\bibitem [{\citenamefont {Renger}\ \emph {et~al.}(2010)\citenamefont {Renger},
  \citenamefont {Kadic}, \citenamefont {Dupont}, \citenamefont
  {A\'{c}imovi\'{c}}, \citenamefont {Guenneau}, \citenamefont {Quidant},\ and\
  \citenamefont {Enoch}}]{Renger_2010}%
  \BibitemOpen
  \bibfield  {author} {\bibinfo {author} {\bibfnamefont {J.}~\bibnamefont
  {Renger}}, \bibinfo {author} {\bibfnamefont {M.}~\bibnamefont {Kadic}},
  \bibinfo {author} {\bibfnamefont {G.}~\bibnamefont {Dupont}}, \bibinfo
  {author} {\bibfnamefont {S.~S.}\ \bibnamefont {A\'{c}imovi\'{c}}}, \bibinfo
  {author} {\bibfnamefont {S.}~\bibnamefont {Guenneau}}, \bibinfo {author}
  {\bibfnamefont {R.}~\bibnamefont {Quidant}}, \ and\ \bibinfo {author}
  {\bibfnamefont {S.}~\bibnamefont {Enoch}},\ }\href@noop {} {\bibfield
  {journal} {\bibinfo  {journal} {Opt. Express}\ }\textbf {\bibinfo {volume}
  {18}},\ \bibinfo {pages} {15757} (\bibinfo {year} {2010})}
\bibitem [{\citenamefont {Chen}\ \emph {et~al.}(2011)\citenamefont {Chen},
  \citenamefont {Luo}, \citenamefont {Zhang}, \citenamefont {Jiang},
  \citenamefont {Pendry},\ and\ \citenamefont {Zhang}}]{chen_2011}%
  \BibitemOpen
  \bibfield  {author} {\bibinfo {author} {\bibfnamefont {X.}~\bibnamefont
  {Chen}}, \bibinfo {author} {\bibfnamefont {Y.}~\bibnamefont {Luo}}, \bibinfo
  {author} {\bibfnamefont {J.}~\bibnamefont {Zhang}}, \bibinfo {author}
  {\bibfnamefont {K.}~\bibnamefont {Jiang}}, \bibinfo {author} {\bibfnamefont
  {J.~B.}\ \bibnamefont {Pendry}}, \ and\ \bibinfo {author} {\bibfnamefont
  {S.}~\bibnamefont {Zhang}},\ }\href@noop {} {\bibfield  {journal} {\bibinfo
  {journal} {Nature Com.}\ }\textbf {\bibinfo {volume} {2}},\ \bibinfo {pages}
  {176} (\bibinfo {year} {2011})} 
\bibitem [{\citenamefont {Schonbrun}\ \emph {et~al.}(2007)\citenamefont
  {Schonbrun}, \citenamefont {Wu}, \citenamefont {Park}, \citenamefont
  {Yamashita}, \citenamefont {Summers}, \citenamefont {Abashin},\ and\
  \citenamefont {Fainman}}]{Schonbrun_2007}%
  \BibitemOpen
  \bibfield  {author} {\bibinfo {author} {\bibfnamefont {E.}~\bibnamefont
  {Schonbrun}}, \bibinfo {author} {\bibfnamefont {Q.}~\bibnamefont {Wu}},
  \bibinfo {author} {\bibfnamefont {W.}~\bibnamefont {Park}}, \bibinfo {author}
  {\bibfnamefont {T.}~\bibnamefont {Yamashita}}, \bibinfo {author}
  {\bibfnamefont {C.~J.}\ \bibnamefont {Summers}}, \bibinfo {author}
  {\bibfnamefont {M.}~\bibnamefont {Abashin}}, \ and\ \bibinfo {author}
  {\bibfnamefont {Y.}~\bibnamefont {Fainman}},\ }\href@noop {} {\bibfield
  {journal} {\bibinfo  {journal} {Appl. Phys. Lett.}\ }\textbf {\bibinfo
  {volume} {90}},\ \bibinfo {pages} {041113} (\bibinfo {year}
  {2007})} 
\bibitem [{\citenamefont {Zolla}\ \emph {et~al.}(2007)\citenamefont {Zolla},
  \citenamefont {Guenneau}, \citenamefont {Nicolet},\ and\ \citenamefont
  {Pendry}}]{Zolla_2007}%
  \BibitemOpen
  \bibfield  {author} {\bibinfo {author} {\bibfnamefont {F.}~\bibnamefont
  {Zolla}}, \bibinfo {author} {\bibfnamefont {S.}~\bibnamefont {Guenneau}},
  \bibinfo {author} {\bibfnamefont {A.}~\bibnamefont {Nicolet}}, \ and\
  \bibinfo {author} {\bibfnamefont {J.~B.}\ \bibnamefont {Pendry}},\
  }\href@noop {} {\bibfield  {journal} {\bibinfo  {journal} {Opt. Lett.}\
  }\textbf {\bibinfo {volume} {32}},\ \bibinfo {pages} {1069} (\bibinfo {year}
  {2007})} 
\bibitem [{\citenamefont {Fabre}\ \emph {et~al.}(2008)\citenamefont {Fabre},
  \citenamefont {Lalouat}, \citenamefont {Cluzel}, \citenamefont {M{\'e}lique},
  \citenamefont {Lippens}, \citenamefont {de~Fornel},\ and\ \citenamefont
  {Vanb{\'e}sien}}]{Fabre_2008}%
  \BibitemOpen
  \bibfield  {author} {\bibinfo {author} {\bibfnamefont {N.}~\bibnamefont
  {Fabre}}, \bibinfo {author} {\bibfnamefont {L.}~\bibnamefont {Lalouat}},
  \bibinfo {author} {\bibfnamefont {B.}~\bibnamefont {Cluzel}}, \bibinfo
  {author} {\bibfnamefont {X.}~\bibnamefont {M{\'e}lique}}, \bibinfo {author}
  {\bibfnamefont {D.}~\bibnamefont {Lippens}}, \bibinfo {author} {\bibfnamefont
  {F.}~\bibnamefont {de~Fornel}}, \ and\ \bibinfo {author} {\bibfnamefont
  {O.}~\bibnamefont {Vanb{\'e}sien}},\ }\href@noop {} {\bibfield  {journal}
  {\bibinfo  {journal} {Phys. Rev. Lett.}\ }\textbf {\bibinfo {volume} {101}},\
  \bibinfo {pages} {073901} (\bibinfo {year} {2008})} 
\bibitem [{\citenamefont {Zhang}\ \emph {et~al.}(2008)\citenamefont {Zhang},
  \citenamefont {Jin},\ and\ \citenamefont {He}}]{Zhang_2008}%
  \BibitemOpen
  \bibfield  {author} {\bibinfo {author} {\bibfnamefont {P.}~\bibnamefont
  {Zhang}}, \bibinfo {author} {\bibfnamefont {Y.}~\bibnamefont {Jin}}, \ and\
  \bibinfo {author} {\bibfnamefont {S.}~\bibnamefont {He}},\ }\href@noop {}
  {\bibfield  {journal} {\bibinfo  {journal} {Opt. Express}\ }\textbf {\bibinfo
  {volume} {16}},\ \bibinfo {pages} {3161} (\bibinfo {year}
  {2008})} 
\bibitem [{\citenamefont {Liao}\ and\ \citenamefont {Liu}(1993)}]{Liao_1993}%
  \BibitemOpen
  \bibfield  {author} {\bibinfo {author} {\bibfnamefont {G.}~\bibnamefont
  {Liao}}\ and\ \bibinfo {author} {\bibfnamefont {H.}~\bibnamefont {Liu}},\
  }\href@noop {} {\bibfield  {journal} {\bibinfo  {journal} {Num. Math. PDEs}\
  }\textbf {\bibinfo {volume} {9}} (\bibinfo {year} {1993})}
\bibitem [{\citenamefont {Liang}\ and\ \citenamefont {Li}(2011)}]{Liang_2011}%
  \BibitemOpen
  \bibfield  {author} {\bibinfo {author} {\bibfnamefont {Z.}~\bibnamefont
  {Liang}}\ and\ \bibinfo {author} {\bibfnamefont {J.}~\bibnamefont {Li}},\
  }\href@noop {} {\bibfield  {journal} {\bibinfo  {journal} {Opt. Express}\
  }\textbf {\bibinfo {volume} {19}} (\bibinfo {year} {2011})}
\bibitem [{\citenamefont {Fabre}\ \emph {et~al.}(2006)\citenamefont {Fabre},
  \citenamefont {Fasquel}, \citenamefont {Legrand}, \citenamefont
  {M{\'e}lique}, \citenamefont {Muller}, \citenamefont {Fran{\c{c}}ois},
  \citenamefont {Vanb{\'e}sien},\ and\ \citenamefont {Lippens}}]{Fabre_2006}%
  \BibitemOpen
  \bibfield  {author} {\bibinfo {author} {\bibfnamefont {N.}~\bibnamefont
  {Fabre}}, \bibinfo {author} {\bibfnamefont {S.}~\bibnamefont {Fasquel}},
  \bibinfo {author} {\bibfnamefont {C.}~\bibnamefont {Legrand}}, \bibinfo
  {author} {\bibfnamefont {X.}~\bibnamefont {M{\'e}lique}}, \bibinfo {author}
  {\bibfnamefont {M.}~\bibnamefont {Muller}}, \bibinfo {author} {\bibfnamefont
  {M.}~\bibnamefont {Fran{\c{c}}ois}}, \bibinfo {author} {\bibfnamefont
  {O.}~\bibnamefont {Vanb{\'e}sien}}, \ and\ \bibinfo {author} {\bibfnamefont
  {D.}~\bibnamefont {Lippens}},\ }\href@noop {} {\bibfield  {journal} {\bibinfo
   {journal} {Opto-Electron. Rev.}\ }\textbf {\bibinfo {volume} {14}},\
  \bibinfo {pages} {225} (\bibinfo {year} {2006})} 
\end{thebibliography}

\providecommand{\noopsort}[1]{}\providecommand{\singleletter}[1]{#1}%

\end{document}